\documentclass[11pt,aps]{revtex4}
\usepackage{epsfig}

\def\lp {\left( }
\def\rp {\right) }
\def\lb {\left[ }
\def\rb {\right] }
\def\lc {\left\{ }
\def\rc {\right\} }

\def\nn {\nonumber}

\def\beq{\begin{equation}}
\def\eeq{\end{equation}}
\def\bea{\begin{eqnarray}}
\def\eea{\end{eqnarray}}
\def\ni{\noindent}

\def\ds {\partial^\mu }

\def\cd {\!\cdot\!}

\def\rar {\rightarrow}

\def\ub {\bar u}
\def\pb {\bar p}

\def\Qs {\not\!\!Q}

\def\st {\tilde{\s}}

\def\M {\th}
\def\Pb {\bar{\Pi}}

\def\Sb {\bar{S}}

\def\mb {M}

\def\sb {\bar{s}}

\def\D {\Delta}

\def\g{\gamma}
\def\D {\Delta}

\def\k {\kappa}

\def\m{\mu}
\def\n{\nu}

\def\p{\pi}
\def\P{\Pi}

\def\s{\sigma}

\def\th {\theta}

\def\cO {{\cal O}}

\def\bfi {\mbox{\boldmath $\phi$}}
\def\bk {\mbox{\boldmath $k$}}
\def\br {\mbox{\boldmath $r$}}

\def\bx {\mbox{\boldmath $x$}}

\def\bnb {\mbox{\boldmath $\nabla$}}

\def\bp {\mbox{\boldmath $p$}}

\def\bpi {\mbox{\boldmath $\pi$}}
\def\btau {\mbox{\boldmath $\tau$}}

\def\bq {\mbox{\boldmath $q$}}

\begin{document}

\title{Hyperon scalar form factors}

\author{C.C. Barros Jr and M.R. Robilotta}

\address{Nuclear Theory and Elementary Particle Phenomenology Group\\
Instituto de F\'{\i}sica, Universidade de S\~{a}o Paulo,\\
C.P. 66318, 05315-970, S\~ao Paulo, SP, Brazil}

\date{\today}

\begin{abstract}
We evaluate the long range part of the hyperon form factors and their contribution to elastic 
pion-hyperon scattering. 
\end{abstract}

\maketitle

\vspace{5mm}

\section{introduction}

High energy proton-nucleus or nucleus-nucleus collisions may release large amounts of energy into very small regions 
and disturb strongly the QCD vacuum.
After hadronization, these reactions produce final states which, typically, contain many pions and a wide 
variety of other particles, including strange ones.
Pion-hyperon $(\p Y)$ interactions are, therefore, one of the many elements that contribute to the detailed 
description of high energy collisions.
The fact that these particles are produced in the same reaction allows one to 
assume that they interact as comovers
in an expanding system, with relative energies which are not very high.

An interesting feature of high energy proton-nucleus collisions is that both hyperons and antihyperons detected in inclusive processes are polarized \cite{YP}.
In 1993, Hama and Kodama \cite{HK} showed that the interaction of hyperons with a background could 
explain the observed polarization.
An optical potential was used and they found out that it had to depend on the particular hyperon considered.
This idea was further developed in a recent work \cite{CCB}, in which the 
background was assumed to be made of pions and the $\p Y$ interactions were 
described by means
of a microscopic chiral model \cite{BH}, adapted from the study of low energy $\p N$ scattering \cite{PiN}.
In this model, the amplitudes are given by tree diagrams, saturated by spin $1/2$ and $3/2$
intermediate states, and supplemented by the exchange of a scalar isoscalar system in the $t$-channel.
This last contribution is related to the scalar form factor of the baryon and 
discussed in the present work.
It is denoted by $\s(t)$, represents the mass density associated with the meson cloud and, at large distances, 
is dominated by triangle diagrams involving the exchange of two pions.

In ref.\cite{BH}, the dependence of the scalar exchange on $t$ was assumed to be of the form $a + bt$ 
and the values of the parameters $a$ and $b$ were found to be important in determining polarization observables.
Moreover, good fits to data could be achieved when these parameters were allowed to scale with the coupling constant 
of the pion to the particular hyperon considered.
These results have motivated the present work, in which we evaluate the pion cloud contribution to the
scalar form factor of spin $1/2$ hyperons and study the behaviour of the parameters $a$ and $b$.
Our calculation follows a procedure used previously in the nucleon case \cite{R}.

The interactions of pions with other hadrons can be well described by means of effective theories,
in which an approximate $SU(2)\times SU(2)$ symmetry is broken by the small pion mass $(\m)$.
In the framework of chiral perturbation theory, the leading term of the nucleon scalar form factor 
is proportional to $\m^2$ and determined directly by the coupling constant $c_1$ of the second order lagrangian \cite{SFF}.
Loop diagrams, which carry the depence on $t$, begin contributing at order $\cO(\m^3)$.
This means that, in a strict calculation, one is not able to predict the value of the parameter $a$, which is 
related with $c_1$.

In order to overcome this difficulty, one notes that, to $\cO(\m^3)$, triangle diagrams are completely determined, 
since they involve only known masses and coupling constants.
To $\cO(\m^4)$, loop diagrams are combined with the constants $c_1, c_2$ and $c_3$.
The study of $\p N$ subthreshold coefficients indicates that $c_1$ is smaller than $c_2$ and $c_3$ and
the values of these last two constants can be explained by means of $\D$ saturation.
This allows the $\cO(\m^4)$ scalar form factor to be well represented by the leading tree contribution 
associated with $c_1$ and and two triangle diagrams, involving $N$ and $\D$ intermediate states, as
in fig.1.

\begin{figure}[hbtp]
\centerline{
\epsfxsize=90.mm
\epsffile{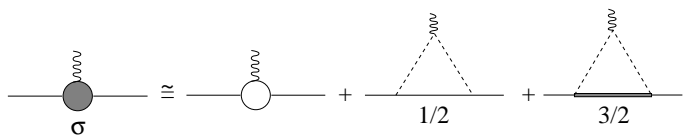}}
\caption{ The scalar form factor (grey blob)  receives contributions from tree
 interactions (white blob)
and triangle diagrams with spin $1/2$ and $3/2$ intermediate states.}
\end{figure}

When one goes to configuration space, these two kinds of contributions split apart in $\st(r)$.
The tree term yields a zero-range $\delta$-function, whereas the triangle diagrams give rise to 
spatially distributed structures, fully determined by known parameters.
As noted in ref.\cite{L}, for the case of hadron polarizabilities, the Fourier transform acts as a filter
which transmits only genuine pion loop effects.
The $\s$-term, defined as $\s\equiv\s(t=0)$, would be given by 

\beq
\s = 4\p \int_0^\infty dr\;r^2\;\st(r)\;.
\label{1.1}
\eeq

The function $\st(r)$ represents the mass density of the nucleon due to its pion cloud.
In non-linear lagrangians, the pion degrees of freedom are better described
by a direction $\hat{\bpi}$ and by an angle $\theta$, embodied into the operator
$U=\exp(i\btau\cdot\hat{\bpi}\;\theta)= \cos\theta + i\btau\cdot\hat{\bpi}\sin\theta$.
In this framework, the value of the dimensional pion field $\bfi=f_\p \;sin\theta \;\hat{\bpi}$ cannot be larger than $f_\p$,
since Goldstone bosons are collective states derived from the $q\bar{q}$ condensate.
Accordingly, a pion cloud corresponds to a transformation of the condensate that surrounds the nucleon  
and its energy density cannot exceed that of the original condensate, whose value is $\m^2 f_\p^2$.
Therefore we define the radius $R$ by the relationship $\st(R)=f_\p^2\m^2$ and write the sigma term as 

\beq
\s = \frac{4}{3} \p R^3 \;f_\p^2 \m^2+ 4\p \int_R^\infty dr\;r^2\;\st(r)\;.
\label{1.2}
\eeq

In the case of the nucleon\cite{R}, this expression yields $\s=46$ MeV, a value quite close to that prescribed in 
ref.\cite{GLS}. 
In the sequence, we extend this procedure to the case of strange baryons.

\section{formalism}

The scalar form factor for a spin $1/2$ baryon B is defined as 
$ <B(p') | - {\cal L}_{sb} | B(p) > \equiv \s(t) \; \ub(\bp') \;u(\bp) $,
where ${\cal L}_{sb}$ is the chiral symmetry breaking lagrangian and $t=(p-p')^2$.
In this work we assume the scalar form factor of strange baryons to be completely dominated
by the long range two-pion processes shown in fig.1.
The relevant interaction Lagrangians are given by

\beq
{\cal L}_{\p BB'} =  \frac{g_A}{2f_\p} \lb \bar{B'}\g_\m \g_5 \;T_a\; B \rb  \ds \phi_a + h.c.\;,
\label{2.1}
\eeq

\ni
and

\beq
{\cal L}_{\pi BR} = \frac{g_A^*}{2 f_\p} \lb \bar{R}_\m\; T_a\;B \rb \ds \phi_a + h.c.  \;,
\label{2.2}
\eeq

\ni
where $B$, $R$ and $\phi$ denote, respectively, spin $1/2$, spin $3/2$ and pion fields, 
$T$ is a matrix that couples baryons into an isospin 1 state,
$f_\p$ is the pion decay constant whereas
$g_A$ and $g_A^*$ are the coupling constants for the processes $\p B\rar B'$ and $\p B\rar R$.
In table 1 we give the values of $g_A$ and $g_A^*$.
The former were obtained by using  $SU(3)$ relations with the results of ref.\cite{Mart} for the $\p\Lambda\Sigma$ vertex as input and one notes that they
 would not change much if the more recent results of ref. \cite{LW} were used.
The latter were taken from Breit-Wigner fits to resonance decay widths.
The spin $3/2$ propagator for a particle of mass $M$ is written as

\beq
iG^{\m\n}(p)  = -i \; \frac{(\not\!\!{p}+M)}{p^2-M^2}\lp g^{\m\n}-\frac{\g^\m\g^\n}{3}
-\frac{\g^\m p^\n}{3M} + \frac{p^\m \g^\n}{3M} - \frac{2 p^\m p^\n}{3M^2}\rp \;.
\label{2.3}
\eeq

\begin{table}[hbt]
\begin{center}
\caption{Axial coupling constants for processes $\p B\rar B'\;(g_A)$ and $\p B\rar R\;(g_A^*)$.}
\begin{tabular} {|c||c|c|c|c|c|c|c|c|}
\hline

	& $N$ & $\Delta$ & $\Lambda$ & $\Lambda^*$ & $\Sigma$ & $\Sigma^*$ & $\Xi$ & $\Xi^*$  \\ 
	& &  &  & (1405) &  & (1385) &  & (1530)  		\\ \hline
$N$	& 1.25 & 2.82 & - &  - &  - & - & - & -		\\ \hline 
$\Lambda$	& - & - & - &  - &  0.98 & 1.74 & - & -		\\ \hline 
$\Sigma$ 	& - & - & 0.98 &  1.63 &  0.52 & $\sim 0$ & - & -		\\ \hline 
$\Xi$	& - & - & - &  - &  - & - & 0.28 & 0.84		\\ \hline

\end{tabular}
\end{center}
\end{table}

The initial and final baryon momenta are denoted by $p$ and $p'$, whereas $k$ and $k'$ are the momenta of the 
exchanged pions.
We also use the variables $P =  (p'+p)/2$, $q = k'-k = p - p'$, $Q = (k+k')/2$, and $\pb=(p+k)=(p'+k')$.
The external baryons are on shell and one has $m^2 =  P^2 + q^2/4$ and $P\cd q = 0$.
This allows one to write  
$t = q^2$ and $Q^2 = (k^2-\m^2)/2 + (k'^2-\m^2)/2 + \lp \m^2 -t / 4 \rp$.

The contribution of an intermediate particle of spin $s$ and mass $M$ to the scalar form factor is given by

\beq
\s_s(t;M)\; \ub\;u = - \m^2 \lp \frac{g_A}{2 f_\p}\rp^2 \lp T_a^\dagger T_a \rp
\int [\cdots] [\ub\; \Lambda_s \;u] \;,
\label{2.4}
\eeq

\ni
with

\beq
\int [\cdots] = \int \frac{d^4Q}{(2\pi)^4}\; \frac{1}{[ (Q\!-\!q/2) ^2\! -\!\m^2] [(Q\!+\!q/2)^2 \!-\!\m^2]} \;,
\label{2.5}
\eeq

\bea
&& [\ub\; \Lambda_{1/2}\;u] = \ub \; \lc -(m\!+\!M) - \frac{(m\!-\!M)(m\!+\!M)^2}{s\!-\!M^2}
+ \lb 1 + \frac{(m\!+\!M)^2}{s\!-\!M^2} \rb \Qs  \rc \;u \;,
\label{2.6}\\[4mm]
&& [\ub\;\Lambda_{3/2} \; u] 
= - \ub \; \lc \lb \frac{1}{s\!-\!M^2}\lp (m\!+\!M) (\m^2-t/2)
-\;\frac{(2 M + m) }{6 M^2}\m^4 \rp
\right.\right.
\nn\\[2mm]
&& \left.\left.
+ \lp \frac{m^2\!-\!M^2}{s\!-\!M^2}-1 \rp \;\frac{(m\!+\!M)}{6 M^2} 
\lp (m\!+\!M) (2 M - m ) + 2\m^2 \rp
-\;\frac{m\; (s\!-\! m^2)}{6 M^2}\rb
\right.
\nn\\[2mm]
&& \left.
+\lb \frac{1}{s\!-\! M^2} \lp (\m^2-t/2)
+\frac{2 m}{3}(m\!+\!M)-\frac{(m\!+\!M)\m^2}{3M} - \frac{\m^4}{6M^2} \rp 
\right.\right.
\nn\\[2mm]
&& \left.\left.
+ \lp \frac{m^2\!-\!M^2}{s\!-\!M^2}-1 \rp \;\frac{1}{6M^2} \lp M^2 + 2m M - m^2 + 2\m^2 \rp  
-\;\frac{s\!-\!m^2}{6M^2} \rb \Qs \rc \;u
\label{2.7}\;.
\eea

In writing these expressions we have used $s = \pb^2 = [ Q^2\! +\! 2 P\cd Q \! -\! t / 4\! +\! m^2 ]$ 
and replaced $k^2$ and $k'^2$ in the numerator with $\m^2$.
This approximation amounts to neglecting short range interactions, since terms proportional to 
$(k^2-\m^2)$ and $(k'^2-\m^2)$ in the numerator may be used to cancel pion propagators in eq.(\ref{2.4}).

Using the loop integrals $\P$ defined in appendix A, we obtain 

\bea
&& \s_{1/2}(t) = \frac{T_a^\dagger T_a}{(4\p)^2}\lp \frac{g_A \m}{2f_\p}\rp^2  (m\!+\!M)
\lb \P_{cc}^{(000)}-\;\frac{m^2\!-\!M^2}{2m\m}\;\P_{\sb c}^{(000)} -\;\frac{m+M}{2m}\;\P_{\sb c}^{(001)} \rb\;,
\label{2.8}\\[4mm]
&& \s_{3/2}(t) = \frac{T_a^\dagger T_a}{(4\p)^2} \lp \frac{g_A^* \m}{2f_\p}\rp^2 \frac{1}{6M^2}
\lc - \lb (m\!+\!M)^2 (2M\!-\!m) + 2 \m^2 (m\!+\!M)
+ (\m^2\!-\!t/2) \; m  \rb \P_{cc}^{(000)}\right.
\nn\\[2mm]
&& \left.
- 2 \m^2 m  \; \Pb_{cc}^{(000)}
+ \lb (m^2\!-\!M^2) (m\!+\!M)^2 (2M\!-\!m) 
+ 2\m^2 (m\!+\!M)(m^2\!-\!M^2) \right.\right.
\nn\\[2mm]
&& \left.\left. + 6  (\m^2\!-\! t/2) M^2 (m\!+\!M) 
-\m^4 (2M\!+\!m) \rb \frac{\P_{\sb c}^{(000)}}{2m\m} 
+ \lb (m\!+\!M)^2 (4mM\!-\!m^2\!-\!M^2) \right.\right.
\nn\\[2mm]
&& \left.\left.
+ 6 M^2 (\m^2\!-\! t/2) 
- 2 \m^2 (m\!+\!M) (2M\!-\!m) 
- \m^4 \rb \frac{\P_{\sb c}^{(001)}}{2m} \rc\;.
\label{2.9}
\eea

These results can be used in numerical evaluations of the scalar form factor.
However, they do not exhibit an explicit chiral structure.
Using the results of appendix A, we express the integrals $\Pb_{cc}^{(000)}$ and 
$\P_{\sb c}^{(001)}$ in terms of $\P_{cc}^{(000)}$ and $\P_{\sb c}^{(000)}$
and obtain

\bea
\s_{1/2}(t) &=& \frac{T_a^\dagger T_a}{(4\p)^2}\lp\frac{g_A \m}{2 f_\p}\rp^2 \frac{(m\!+\!M)}{8m^3}
\lc \lb 4\, (m\!-\!M)\,m^2 -t\,(m\!+\!M) \rb \P_{cc}^{(000)}\right.
\nn\\[2mm]
&+& \left. \lb - 4\, (m\!-\!M)\,m^2 + t\, (m\!+\!M)
+ 4m^2 (\m^2\!-\!t/2)/(m\!-\!M) \rb\frac{m^2\!-\!M^2}{2m\m} \P_{\sb c}^{(000)} \rc \;,
\label{2.10}\\[4mm]
\s_{3/2}(t)  &=&  \frac{T_a^\dagger T_a}{(4\p)^2}\lp \frac{g_A^* \m}{2 f_\p}\rp ^2 \frac{(m\!+\!M)}{24 m M^2} 
\lc \lb 2\,(m^2\!-\!M^2)(m\!+\!M) \right.\right.
\nn\\[2mm]
&& \left.\left.
+\frac{(\m^2-t/2)}{m^2(m\!+\!M)}
\lp \!M^4\!-\!2mM^3\!+\!6m^2M^2\!-\!2m^3M\!-13m^4/3\rp \right.\right.
\nn\\[2mm]
&& \left.\left. +\frac{\m^2}{m^2}
\lp -\!M^3\!+\!3mM^2\!-\!5m^2M\!-\!5m^3\!- 4m^4/3(m\!+\!M)\rp 
\rb \P_{cc}^{(000)}\right.
\nn\\[2mm]
&+& \left. \lb - 2 (m^2\!-\!M^2)(m\!+\!M) 
+ 8 (\m^2-t/2) m M^2 /(m^2\!-\!M^2) \right.\right.
\nn\\[2mm]
&& \left.\left.
+\frac{(\m^2-t/2)}{m^2(m\!+\!M)}
\lp -\! M^4\!+\!2mM^3\!-\!8m^2M^2\!-\!2m^3M\!+\!m^4 \rp \right.\right.
\nn\\[2mm]
&& \left.\left. + \frac{\m^2}{m^2}
\lp M^3\!-\!3mM^2\!+\!5m^2M\!+\!5m^3 \rp \rb 
\frac{(m^2\!-\!M^2)}{2m\m} \P_{\sb c}^{(000)} \rc\;.
\label{2.11}
\eea

For $M=m$, one has $\P_{cc}^{(000)}\sim \P_{\sb c}^{(000)}\sim  \cO(\m^0)$ and hence 
the loop contribution to $\s(t)$ is $\cO(\m^3)$, as expected.
This result does not change when $M\neq m$ because, from eq.(\ref{a11}), one 
learns that
$[\P_{cc}^{(000)} - (m^2-\mb^2)\;\P_{\sb c}^{(000)}/2m\m] \sim \cO(\m^3)$.

The scalar form factor in configuration space is obtained by going to the Breit frame and writing

\beq
\st(r) = \int \frac{d\bq}{(2\p)^3}\; e^{-i \bq \cdot \br}\; \s(t)
\label{2.12}
\eeq 

\ni
and the corresponding expressions are derived from eqs.(\ref{2.10}) and (\ref{2.11}) trough the replacements 
$t\rar \bnb^2$ and $\P\rar \m^3 S$, with  $S$ given by eqs.(\ref{a13}-\ref{a15}).

\section{results and conclusions}

\begin{figure}[hbtp]
\centerline{
\epsfxsize=70.mm
\epsffile{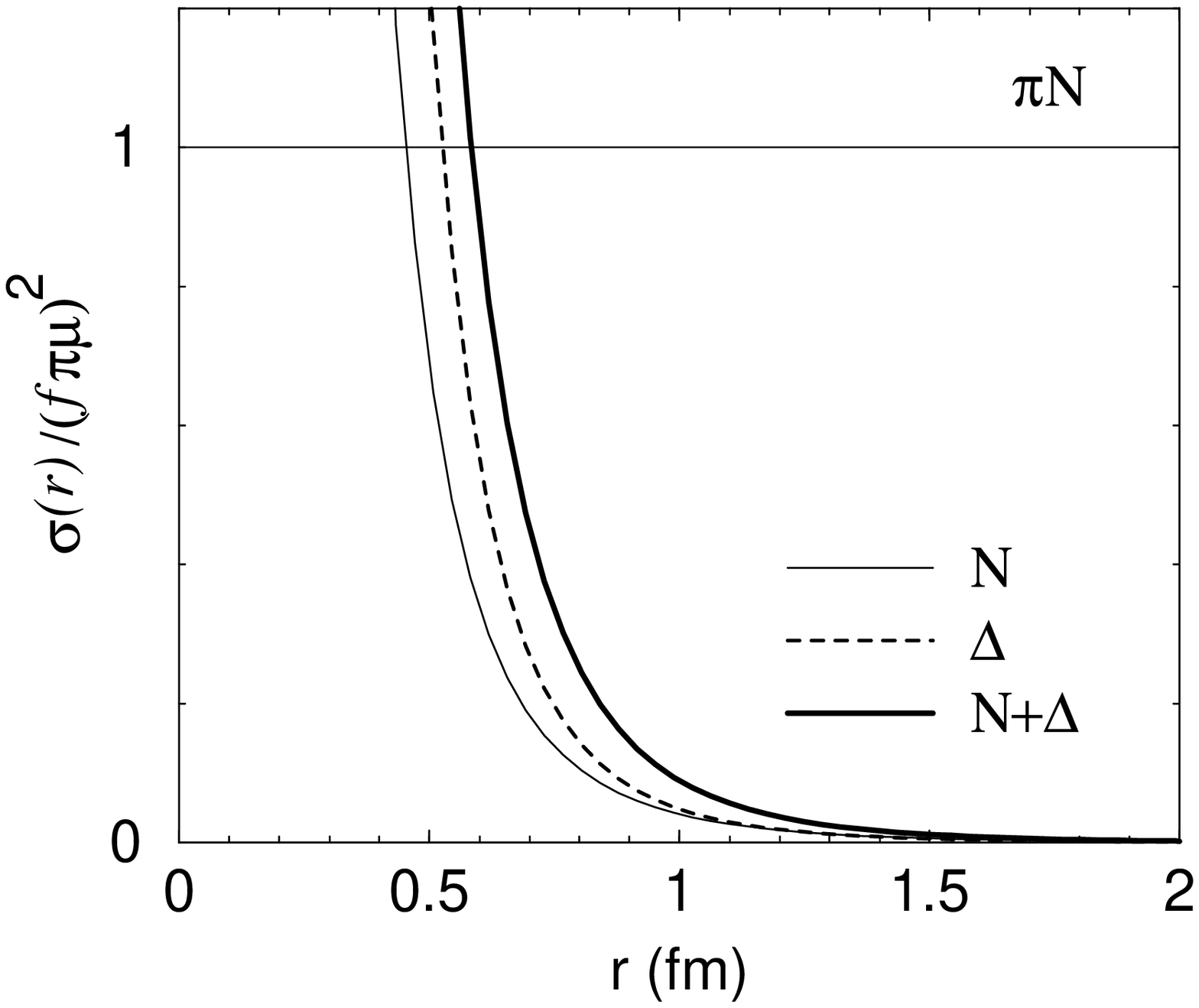}
\epsfxsize=70.mm
\epsffile{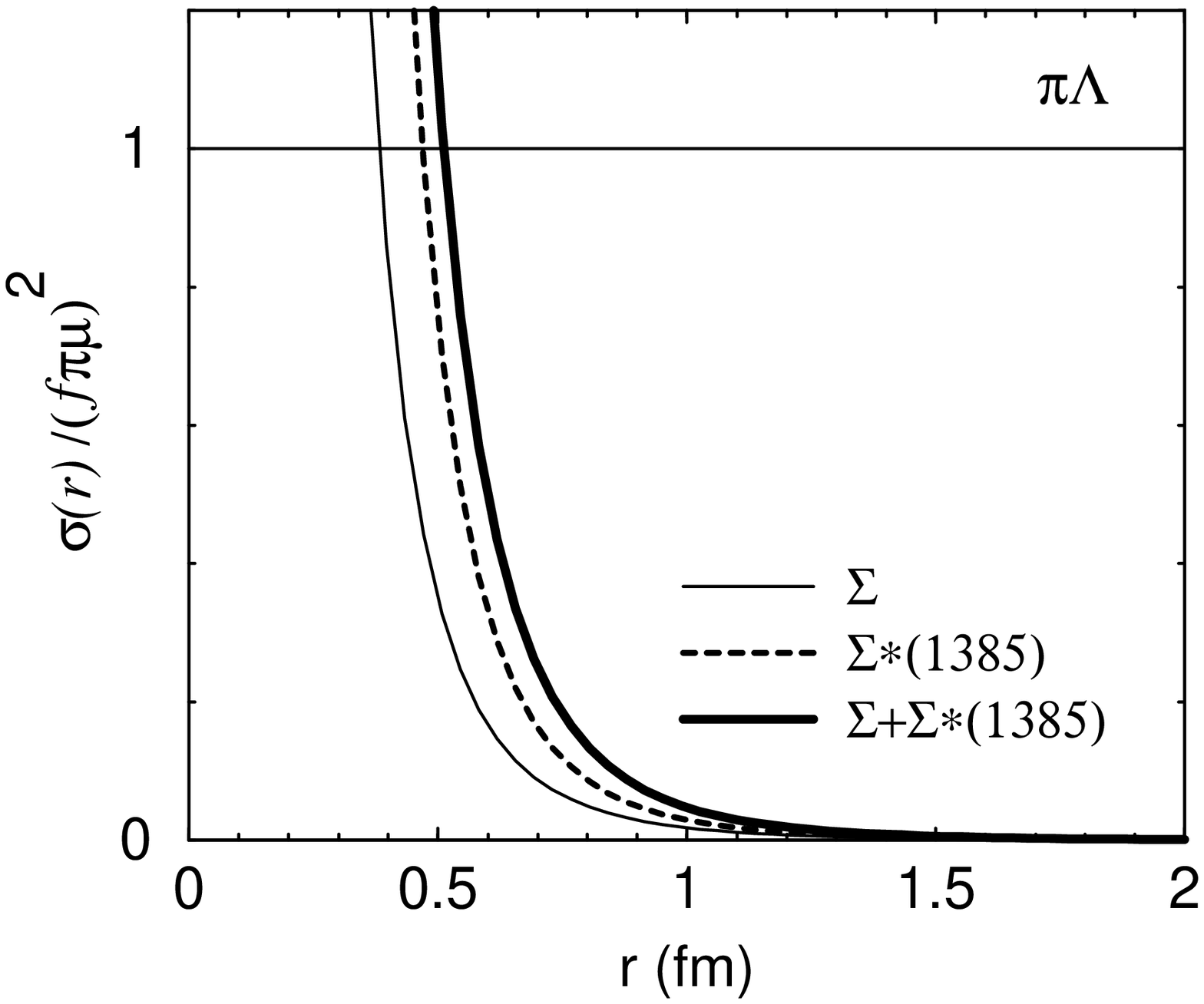}}
\centerline{
\epsfxsize=70.mm
\epsffile{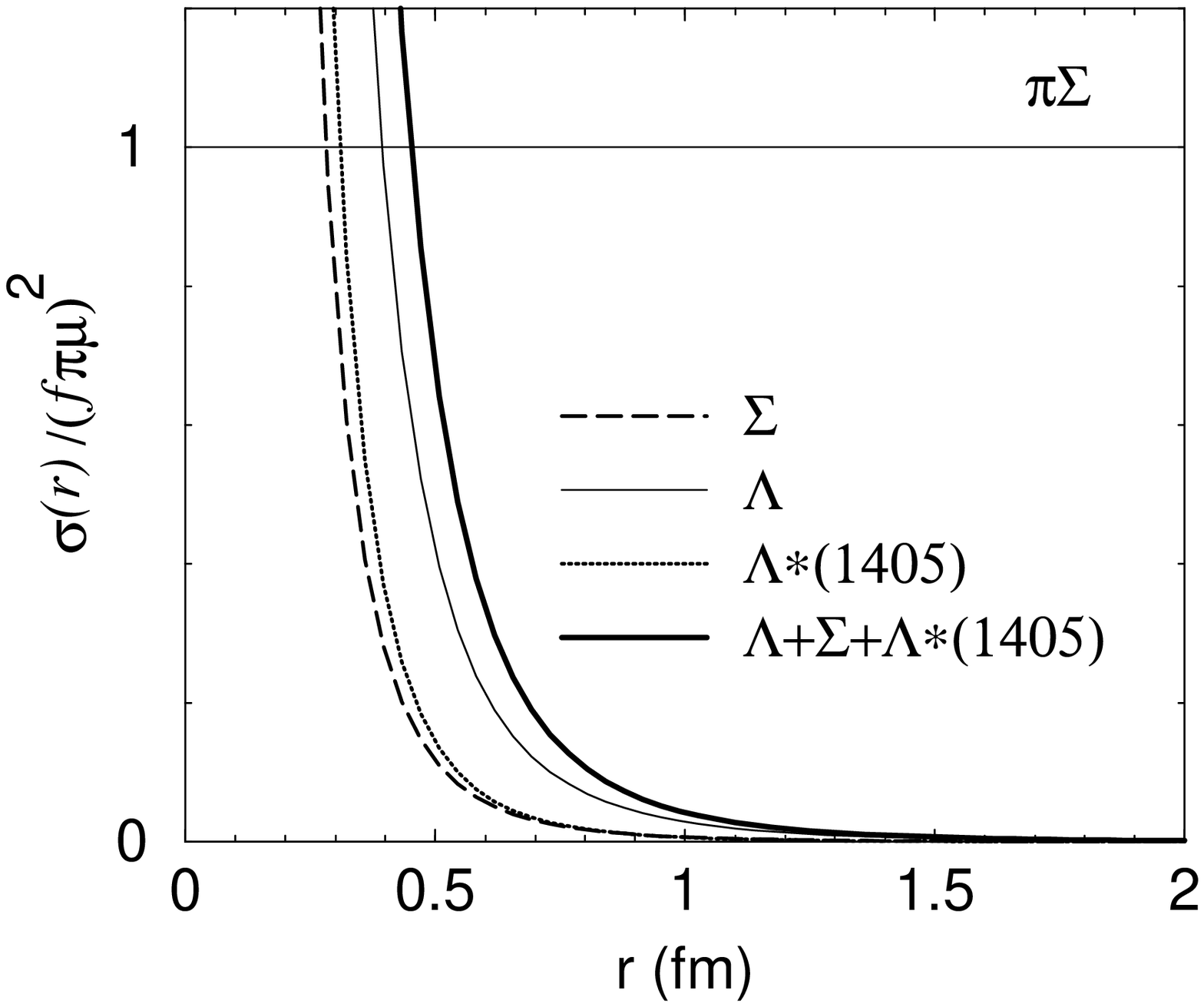}
\epsfxsize=70.mm
\epsffile{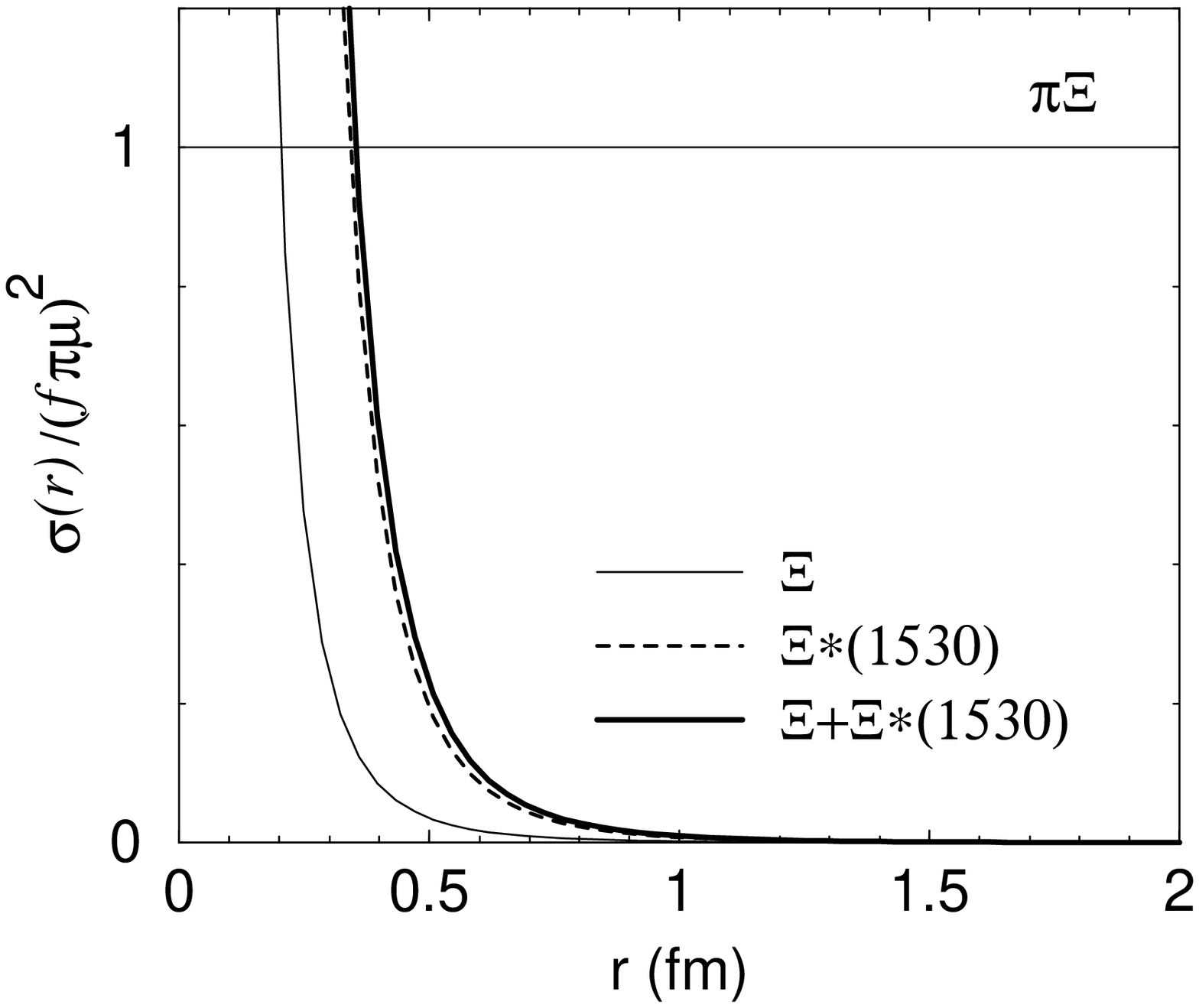}}
\caption{Intermediate state contributions to the scalar form factor of the spin $1/2$ baryons.}
\end{figure}

We begin by discussing our results for the ratios $\st(r)/f_\p^2\m^2$ for the spin $1/2$ baryons.
In fig.2 we display the individual contributions of the various intermediate states as functions of the distance
and one notes that the role of resonances is rather important.
Comparing this feature with the fact that, in the framework of chiral symmetry, one has $\s_{1/2}\rar \cO(\m^3)$ and $\s_{3/2}\rar \cO(\m^4)$, one learns that the power counting hierarchy is subverted around $r \sim 1$fm.

\begin{figure}[hbtp]
\centerline{
\epsfxsize=80.mm
\epsffile{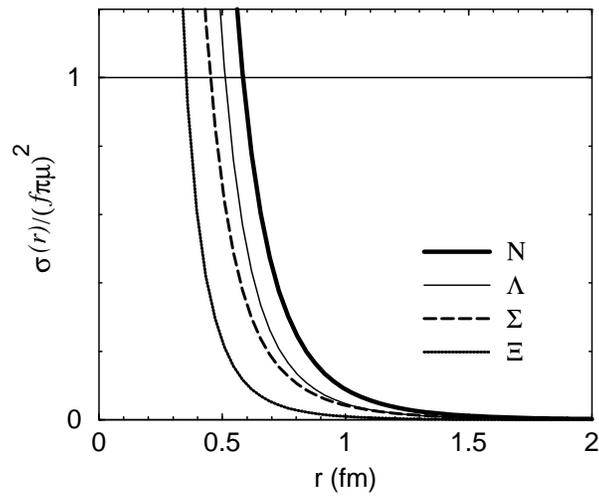}}
\caption{ Full results for the scalar form factor of the spin $1/2$ 
baryons.}
\end{figure}

The full curves for the $N$, $\Lambda$, $\Sigma$ and $\Xi$ states are shown in fig.3.
The values of the distance $R$ for which $\st(R)/f_\p^2\m^2=1$ and of the the $\s$-term, 
calculated by means of eq.({\ref{1.2}), are given in table 2.
One finds that heavier systems correspond to smaller values of these quantities, 
but one should bear in mind that the coupling constants of table 1 also intervene.

\begin{table}[hbt]
\begin{center}
\caption{Baryon radius $(R)$, $\s$-term$=\s(0)$ and $\s(2\m^2)$.}
\begin{tabular} {|c|c|c|c|c|}
\hline

		& $N$ 	& $\Lambda$ 	& $\Sigma$ 	& $\Xi$ 		\\ \hline
$R$ (fm)		& 0.58 	& 0.51		& 0.45		&  0.35 		\\ \hline 
$\s$ (MeV)	& 46.0 	& 33.5		& 29.2		&  12.0 		\\ \hline 
$\s(2\m^2)$ (MeV)	& 57.6 	& 39.3		& 36.2		&  13.25		\\ \hline 

\end{tabular}
\end{center}
\end{table}

These results allow the scalar form factor in momentum space to be written,
for each of the baryons considered, as

\beq
\s(t) = \s + \sum \lb \s_{1/2}(t) -  \s_{1/2}(0) + \s_{3/2}(t) -  \s_{3/2}(0) \rb \;,
\label{3.1}
\eeq

\ni
where the summation runs over possible intermediate states.
In table 2 we also quote $\s(2\m^2)$, the value of this function at the Cheng-Dashen point.  
The contribution of this function to the isospin even subamplitude $A^+$ of the $\p Y$ scattering amplitude \cite{BH} is 
given by $ \s(t)/f_\p^2$.
It corresponds to the $t$-channel exchange of a scalar system in elastic scattering and, with the motivation
discussed in the introduction, we write

\beq
\frac{\s(t)}{f_\p^2} = a + b\, t + c\, t^2\;.
\label{3.2}
\eeq

With the coefficients given in table 3, this series represents well the calculated function for momenta up to $450$ MeV.
These results support qualitatively the assumptions made in ref.\cite{BH}, namely that the hyperon parameters
must be smaller than those o f the nucleon. 
On the other hand, the values used in ref.\cite{CCB}, also given in table 3, must be updated and
a full calculation of the hyperon polarization in high energy proton-nucleus collisions based on the theoretical values 
will be reported elsewhere.

\begin{table}[hbt]
\begin{center}
\caption{Coefficients of the series given in eq.(\ref{3.2}) and values used in ref.\cite{CCB}, within brackets.}
\begin{tabular} {|c|c|c|c|c|}
\hline

		& $N$	& $\Lambda$	& $\Sigma$ 	& $\Xi$ 	\\ \hline
$a(\m^{-1})$	&   0.7423 [0.25] &  0.5390 [0.22]	& 0.4698	[0.13]	&  0.1936 [0.07]	\\ \hline 
$b(\m^{-3})$	&  0.0690	[0.40] & 0.0361  [0.35]	& 0.0322	[0.20]	&  0.0074	[0.12]\\ \hline
$c(\m^{-5})$	&  0.0015	[0]      & 0.0006  [0]		& 0.0011	[0]	& 0.0001 [0]	\\ \hline

\end{tabular}
\end{center}
\end{table}

\begin{acknowledgments}
This work was supported by FAPESP (brazilian agency).
\end{acknowledgments}

\appendix\section{loop integrals}

The basic loop integrals needed in this work are given by

\bea
&& I_{cc}^{\m\cdots} =  \int [\cdots] \; \lp\frac{Q^\m}{\m}\cdots \rp \;,
\label{a1}\\
&& I_{\sb c}^{\m\cdots}= \int [\cdots] \; \lp  \frac{Q^\m}{\m}\cdots \rp \;
\frac{2m\m}{ [s -\mb^2 ] } \;.
\label{a2}
\eea

All denominators are symmetric under $q \rightarrow - q$ and hence results cannot contain odd powers of this variable. 
The integrals are dimensionless and have the following tensor structure

\bea
&& I_{cc} =  \frac{i}{(4\pi)^2} \lc \P_{cc}^{(000)}\rc \;,
\label{a3}\\
&& I_{cc}^{\m\n} = \frac{i}{(4\pi)^2}\lc \frac{q^\m q^\n}{\m^2} \; \P_{cc}^{(200)} 
+ g^{\m\n}\;\Pb_{cc}^{(000)}\rc \;,
\label{a4}\\
&& I_{\sb c} =  \frac{i}{(4\pi)^2}\lc \P_{\sb c}^{(000)} \rc\;,
\label{a5}\\
&& I_{\sb c}^{\m} =  \frac{i}{(4\pi)^2} \lc \frac{P^\m}{m} \; \P_{\sb c}^{(001)}\rc \;.
\label{a6}
\eea

The usual Feynman techniques for loop integration allow us to write

\bea
&& \P_{cc}^{(n00)} = - \int_0^1d a\;  (1/2-a)^n\; \ln \lp \frac{D_{cc}}{\m^2}\rp  \;,
\label{a7}\\[2mm]
&& \Pb_{cc}^{(000)}= - \;\frac{1}{2} \int_0^1 d a \; \frac{D_{cc}}{\m^2} \;
\ln \lp \frac{D_{cc}}{\m^2}\rp  \;,
\label{a8}\\[2mm]
&& \P_{\sb c}^{(00n)}=  \lp\!- 2m /\m \rp^{n+1} \int_0^1 d a\;a \int_0^1 d b \; 
\frac{\m^2\; (ab/2)^n}{D_{\sb c}}\;,
\label{a9}
\eea

\ni
with

\bea
&& D_{cc} = -a(1-a)\;q^2 + \m^2 \;,
\nn\\
&& D_{\sb c} = -a(1-a)(1-b)\;q^2 + [\m^2 -ab\;(\m^2+m^2-\mb^2) + a^2b^2 \; m^2] \;.
\nn
\eea

We now derive two relationships among integrals that allow results to be simplified.
Using eq.(\ref{a1}), we have

\bea
&& \frac{q^\m}{\m} \; I_{cc}^{\m\n} = \int [...]\; \frac{Q^\n Q\cd q}{\m^3}
= \int [...]\;  \frac{Q^\n}{2\m^3}\lb (k'^2\!-\!\m^2) - (k^2\!-\!\m^2) \rb
= \cdots 
\nn\\[2mm]
&& g_{\m\n} \;  I_{cc}^{\m\n} =  \int [...]\; \frac{Q^2}{\m^2}
= \int [...] \lb \lp 1\! -\!t/ 4\m^2 \rp + (k^2-\m^2)/\m^2 \rb 
= \lp 1\!-\! t /4\m^2 \rp I_{cc} + \cdots 
\nn
\eea

\ni
where the ellipsis indicate the contributions of short range terms.
These results and eq.(\ref{a4}) yield 

\beq
\bullet \;\;\Pb_{cc}^{(000)} = \frac{1}{3}\lp 1-\frac{t}{4\m^2}\rp \P_{cc}^{(000)} + \cdots \;.
\label{a10}
\eeq

From eq.(\ref{a2}) we get 

\bea
\frac{P_\m}{m} \; I_{\sb c}^\m 
&=& \int [...]\; \frac{P\cd Q}{(s\!-\!\mb^2)}
= \int [...] \lb 1 - \frac{(Q^2\!-\!q^2/4\!+\!m^2\!-\!\mb^2)}{(s\!-\!\mb^2)}\rb 
\nn\\[2mm]
&=&  I_{cc}-\lb \lp\m^2 - t/ 2\rp+ (m^2-\mb^2) \rb\frac{1}{2m\m}\; I_{\sb c}+ \cdots \;. 
\nn
\eea

Definitions (\ref{a3}-\ref{a6}) then give rise to 

\beq
\bullet \;\; \frac{P^2}{m^2} \;\P_{\sb c}^{(001)}
= \P_{cc}^{(000)} - \lb \lp \m^2 - t/2\rp + (m^2-\mb^2) \rb \frac{1}{2m\m}\;\P_{\sb c}^{(000)} +\cdots
\label{a11}
\eeq

The dimensionless configuration space functions $S$ are defined  as

\beq
S = \int \frac{d\bk}{(2\p)^3}\; e^{-i\bk \cdot \bx}\; \P \;.
\label{a12}
\eeq 

\ni
with $x=\m r$ and $\bk=\bq/\m$.
The results 

\bea
&& \int\frac{d\bk}{(2\p)^3}\; e^{-i\bk\cdot\bx}\; \ln\lp 1+ \frac{\bk^2}{\M^2}\rp  
= -\;\frac{2\M}{4\p} \lp 1 +\frac{1}{\M x} \rp \frac{e^{-\M x}}{x^2}\;,
\nn\\[2mm]
&& \int\frac{d\bk}{(2\p)^3}\; e^{-i\bk\cdot\bx}\;(\k^2)\; \ln\lp 1+ \frac{\bk^2}{\M^2}\rp  
= \frac{2\M^3}{4\p} \lp 1 + \frac{3}{\M x} + \frac{6}{\M^2 x^2} + \frac{6}{\M^3 x^3} \rp \frac{e^{-\M x}}{x^2}\;,
\nn\\[2mm]
&& \int\frac{d\bk}{(2\p)^3}\; e^{-i\bk\cdot\bx}\; \frac{1}{(\bk^2+\M^2)} 
= \frac{1}{4\p}\;\frac{e^{-\M x}}{x}\;,
\nn
\eea

\ni
produce

\bea
&& S_{cc}^{(000)} =  \frac{2}{4\p}  
\int_0^1 d a \; \th_{cc}\lp 1 + \frac{1}{x\th_{cc}} \rp \frac{e^{-x\th_{cc}}}{x^2}
=  \frac{4}{4\p} \frac{K_1(2x)}{x^2} \;,
\label{a13}\\[2mm]
&& \Sb_{cc}^{(000)} = -\; \frac{2}{4\p} 
\int_0^1 d a \; \lp 1 + \frac{3}{x\th_{cc}} +\frac{3}{x^2\th_{cc}^2} \rp 
\frac{e^{-x \th_{cc}}}{x^3} 
= -\; \frac{2}{4\p}\lb \frac{K_0(2x)}{x^3} + \frac{K_1(2x)}{x^4} \rb\;,
\label{a14}\\[2mm]
&& S_{\sb c}^{(00n)} =  \frac{\lp- 2m/\m\rp^{n+1}}{4\p}
\int_0^1 d a \;a \int_0^1 d b \;\frac{(ab/2)^n}{a(1-a)(1-b)} \; \frac{e^{-x \th_{\sb c}}}{x} \;,
\label{a15}
\eea

\ni
where

\bea
&&\th_{cc}^2 = 1/a(1-a) \;,
\nn\\
&&\th_{\sb c}^2 = \lb 1-ab[1+(m^2-M^2)/\m^2] + a^2b^2 (m/\m)^2 \rb /a(1-a)(1-b) \;.
\nn
\eea


\end{document}